\documentclass{article}

\usepackage{amsmath}

\usepackage{listings}
\usepackage{color}

\definecolor{dkgreen}{rgb}{0,0.6,0}
\definecolor{gray}{rgb}{0.5,0.5,0.5}
\definecolor{mauve}{rgb}{0.58,0,0.82}

\usepackage{graphicx}
\usepackage{color}
\usepackage{pgfplots}
\usepackage{filecontents} 
\usetikzlibrary{patterns}
\usepackage{helvet}
\usepackage[eulergreek]{sansmath}

\pgfplotsset{
tick label style = {font=\sansmath\sffamily\scriptsize},
every axis label/.append style={font=\sffamily\footnotesize\scriptsize},
}
\pgfplotsset{compat=1.3}

\begin{document}

\setlength{\pdfpageheight}{\paperheight}
\setlength{\pdfpagewidth}{\paperwidth}

\title{IPMACC: Open Source OpenACC to CUDA/OpenCL Translator}
%\subtitle{Subtitle Text, if any}
\author{Ahmad Lashgar\\ University of Victoria
		 \and Alireza Majidi \\ Texas A\&M University
        \and Amirali Baniasadi \\ University of Victoria}

\maketitle

\begin{abstract}
In this paper we introduce IPMACC, a framework for translating OpenACC applications to CUDA or OpenCL. IPMACC is composed of set of translators translating OpenACC for C applications to CUDA or OpenCL. The framework uses the system compiler (e.g. nvcc) for generating final accelerator's binary. The framework can be used for extending the OpenACC API, executing OpenACC applications, or obtaining CUDA or OpenCL code which is equivalent to OpenACC code. We verify correctness of our framework under several benchmarks included from Rodinia Benchmark Suit and CUDA SDK. We also compare the performance of CUDA version of the benchmarks to OpenACC version which is compiled by our framework. By comparing CUDA and OpenACC versions, we discuss the limitations of OpenACC in achieving a performance near to highly-optimized CUDA version.
\end{abstract}

\section{Background}
\subsection{CUDA Model}
In CUDA \cite{Nickolls2008}, an application is composed of host and device codes. The host code executes on CPU and the device code executes on system's accelerator. The host controls the operations of the device through procedure calls to CUDA API. CUDA allows programmers to explicitly allocate memory on device and transfer data between the host and the device. The device obtains the device code from kernel and executes it by thousands of light-weight threads, in SIMT style \cite{Lindholm2008}. All threads share common off-chip DRAM memory or global memory. In software, threads are grouped into coarser scheduling elements, referred to as the thread block. Threads within the same block execute concurrently and communicate through a fast, per-block, on-chip software-managed cache, referred to as shared memory. Shared memory is much faster than global memory; e.g. under GTX 280, the latency of global memory and shared memory are 440 and 38 core cycles, respectively \cite{Wong2010}.
\subsection{OpenACC Model}
OpenACC API introduces a set of compiler directives, library routines, and environment variables to offload a region of code from the CPU to the system's accelerator \cite{OpenACC2013}. We refer to this region as the accelerator region. OpenACC has two classes of directives: i) data management and ii) parallelism control. Each directive has clauses providing finer-grain control. Data management directives perform data allocation on the accelerator, data transfer between the host and the accelerator, and passing pointers to the accelerator. Parallelism control directives allow the programmer to mark regions of code, usually work-sharing loops, intended to run in parallel on the accelerator. They also control parallelism granularity, variable sharing/privatization, and variable reduction. OpenACC introduces four levels of parallelism: gang, worker, vector, and thread. In CUDA terminology, these terms best map to kernel, thread block, warp, and thread, respectively.

\begin{lstlisting}[float,caption=OpenACC and CUDA matrix-matrix multiplications., label=matmulc,basicstyle=\scriptsize]
#pragma acc data copy(a[0:LEN*LEN],b[0:LEN*LEN],c[0:LEN*LEN]) 
#pragma acc kernels 
#pragma acc loop independent   
for(i=0; i<LEN; ++i){
#pragma acc loop independent
 for(j=0; j<LEN; ++j){ 
  float sum=0; 
  for(l=0; l<LEN; ++l) sum += a[i*LEN+l]*b[l*LEN+j];
  c[i*LEN+j]=sum; 
 }
} 
                          (a) OpenACC.

__global__
void matrixMul(int *a, int *b, int *c, int len){
 int i=threadIdx.x+blockIdx.x*blockDim.x; 
 int j=threadIdx.y+blockIdx.y*blockDim.y;
 for(int l=0; l<len; ++l) sum=a[i*len+l]*b[l*len+j];
 c[i*len+j]=sum;
}
int main(){
 ...
 bytes=LEN*LEN*sizeof(int);
 cudaMalloc(&a_d, bytes);
 cudaMalloc(&b_d, bytes);
 cudaMalloc(&c_d, bytes);
 cudaMemcpy(a_d, a, bytes, cudaMemcpyHostToDevice);
 cudaMemcpy(b_d, b, bytes, cudaMemcpyHostToDevice);
 dim3 gridSize(LEN/16,LEN/16), blockSize(16,16);
 matrixMul<<<gridSize,blockSize>>>(a_d,b_d,c_d,LEN);
 cudaMemcpy(c, c_d, bytes, cudaMemcpyDeviceToHost);
 ...
}
                            (b) CUDA.
\end{lstlisting}

\subsection{Matrix-Matrix Multiplication Example}
Listing 1a and 1b illustrate a simple matrix-matrix multiplication in OpenACC and CUDA, respectively. Ignoring the directive lines, Listing 1a shows the baseline serial multiplication of a and b, storing the result in c. Each matrix is LEN*LEN in size. The outer loops iterated by i and j variables can be executed in parallel. Listing 1a shows how these loops can be parallelized using OpenACC. In this code, \textit{kernels} directive marks a region intended to be executed on the accelerator. \textit{loop} directive guides the compiler to consider the loop as a parallel work-sharing loop. Programmers can control the parallelism using kernels and loop directives. As an example of parallelism control, the independent clause is used to force the compiler to parallelize the loop. This clause overwrites the compiler's auto-vectorization and loop dependency checking. In Listing 1a, data clauses hint the compiler to copy a, b, and c arrays from the host to the accelerator, and copy them out from the accelerator to the host. For each array, the [start:n] pair indicates that n elements should be copied from the start element of array. ). Listing 1b shows how the parallelization can be exploited in CUDA. global indicates the declaration of kernel code. Parallel threads execute the kernel and operate on different matrix elements, based on their unique indexes (i and j). Inside the host code, device memory is allocated for a, b, and c, keeping the pointer in a\_d, b\_d, and c\_d, respectively. Then, input matrices are copied into device memory. Then, a total of LEN*LEN light-weight accelerator threads are launched on the device to execute matrixMul kernel. After kernel completion, the resulting matrix c d is copied back to the host memory. As presented in Listing 1, OpenACC significantly reduces the accelerator programming effort in comparison to CUDA. OpenACC hides low-level accelerator-related code from the programmer and provides a unified view over both host and accelerator code.

\section{IPMACC Infrastructure}
\begin{figure}[h!]
  \centering
    \includegraphics[width=0.95\textwidth]{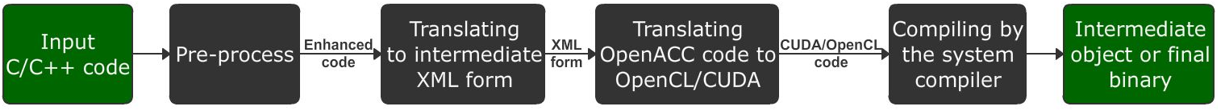}
  \caption{IPMACC compilation pipeline.}\label{pipeline}
\end{figure}

IPMACC compilation process starts with an input C/C++ code which is enhanced by OpenACC API to take advantage of accelerators. The output of the process can be an object code, binary, or C/C++ source code targeted for either CUDA- or OpenCL-capable accelerator. Figure \ref{pipeline} shows the diagram of compilation process which consists of four major stages. In this section, we describe the compilation process in more details. Descriptions in this section can be used to modify IPMACC to generate customized code.

\subsection{Stage 1: Pre-process}
This stage performs pre-processing to normalize/verify the syntax of input C/C++ and OpenACC API. We use uncrustify \cite{Unc2014} to validate C/C++ syntax and normalize its notation. For example, the region of control (if) and loop (while, for) statements will be fully-bracketed. Such polishing passes mark the scope associated with each OpenACC region; simplifying subsequent stages. We have also developed a set of scanners to validate OpenACC API syntax. The scanners assert the validity of directive in respect to OpenACC API and assert the validity of clauses in respect to directive. It also asserts OpenACC restrictions of using nested directives.

\subsection{Stage 2: Translating to Intermediate Form}
This stage transforms the amended input code to intermediate XML form. The XML form has only three types of tags: C code, OpenACC pragma, and for loop. During compilations, the codes encompassed in the C code tags remain unmodified. OpenACC pragma tags will be replaced by proper calls to implement the accelerator-oriented operations. for loop tags will be either parallelized on accelerator or stay as they are (e.g. serial). This decision is made based on the preceding OpenACC directive (e.g. loop directive) or auto-vectorization optimizations. This intermediate representation separates the host/accelerator and remarkably facilitates OpenACC translation in next stage.

\subsection{Stage 3: Translating OpenACC API to target code (CUDA/OpenCL)}
There are nine sequential steps to translate the intermediate XML form to the final CUDA/OpenCL source code.

\subsubsection{Extracting OpenACC regions and retrieving the reversion of C/C++}
This step returns the XML form to C/C++ form while replacing accelerator-related codes (OpenACC pragma tags) with a dummy function call. Meanwhile, the process maintains the OpenACC information related to dummy functions. These are essences used to generate corresponding CUDA/OpenCL codes.

The process conceptually splits the code into two codeblocks (while they are already linked through dummy functions): i) the code bounded by OpenACC API (referred to as Regions code), and ii) the code placed outside of OpenACC boundaries (referred to as Original code). Original code is executed by host. Regions code, which includes OpenACC data clauses and kernels regions, is executed either by host or accelerator.

In this terminology, at this step, each Regions code is replaced by a dummy function call in Original code generating flat host code plus a number of dummy function calls. At the end of compilation, dummy functions are replaced by the target-accelerator codes launching computations on the target accelerator, controlling memory transfers between accelerator and host, and synchronizing host-accelerator operations.

\subsubsection{Retrieve AST of C/C++ code}
This step calculates the abstract syntax tree (AST) of Original code. We use AST representation to find further information about variables, types, and functions which are used in the Regions code but are not declared in that scope. Particularly, the operation searches for the type of variables, size of arrays, and declaration of user-defined non-standard functions/types.

\subsubsection{Find the scope of Regions code (parallel to 2)}
The scope of Regions code contains the declarations/prototypes (function, variable, or type) which are used in the region. It consists of global scope in addition to the scope where the Regions code is called. Since a dummy function call is the representative of a Region code, the scope of dummy function's parent is the scope of declarations/prototypes used in the Regions code. Accordingly, this step finds the parent function calling each dummy call (notice that dummy functions are unique and have only one parent).  The global scope and the scope of parent function are searched for the declarations/prototypes that are referred to in the Regions code.

\subsubsection{Construct kernel code}
This stage constructs the body of kernel -- the targeted code to be executed on the accelerator. This construction includes: i) specifying the available parallelism, ii) sharing loop iterations between concurrent accelerator threads, iii) performing variable reductions, iv) regenerating out-defined declarations/prototypes, and v) specifying kernel arguments.

\subsubsection{Replacing the dummy OpenACC data clause (parallel to 4)}
The dummy function calls which correspond to OpenACC memory management clauses are replaced by OpenACC function calls that implement the data management operations (Table 1). Data managements include host-accelerator pointer exchange, data copy in/out to/from accelerator, and memory allocation. Memory allocation essentially needs the size of memory. The size is either provided by the programmer manually (through the clauses parameters) or detected by the compiler automatically (only the fixed-size arrays are identifiable).

\subsubsection{Finding the declaration of functions/types referred in Regions code (parallel to 4)}
This step finds the non-standard types and non-built-in functions which are called in the Region code. Subsequently, it searches the AST of Original code, which is generated on the 2nd step, to find the declarations of these user-defined non-standard functions/types. Later, these declarations will be appended to the final kernel code.

\subsubsection{Replacing the dummy OpenACC kernels region calls}
This step replaces each dummy function associated with kernel calls with a codeblock performing kernel body setup, kernel argument arrangement, and kernel invocation. There is an extra code in case of variable reduction (e.g. reduction clause in loop directive) that merges results across different thread blocks. We implement variable reduction according to the algorithm proposed in \cite{Hariss2006}.

\subsubsection{Append forward declaration to the code (parallel to 7)}
At this step, the Original code has been enhanced to launch computation on the target accelerator (CUDA/OpenCL -capable accelerator).

\subsubsection{Store the code into the disk with the proper suffix}
This step stores the enhanced Original code on disk, in the same path as the input C/C++ file with \_ipmacc.[cu/c/cpp] suffix.

\subsection{Stage 4: Generating the final object code/binary}
This stage invokes system compiler (nvcc for CUDA backend and g++ in other cases) to generate the target binary. Input to this stage is the source code which is generated in stage 3 and the output can be an object code or executable binary. Operations of this stage are controlled by the ipmacc compilation flags.

\section{Methodology}
{\bf Benchmarks.} We use benchmarks from NVIDIDA CUDA SDK \cite{NVIDIA2014a} and Rodinia benchmark suit \cite{Che2009}. NVIDIA CUDA SDK includes a large set of CUDA test-cases, each implementing a massively-parallel body of an application in CUDA very efficiently. Each test-case also includes a serial C/C++ implementation. We developed an OpenACC version of these benchmarks over the serial C/C++ code. Rodinia is a GPGPU benchmark suite composed of a wide set of workloads implemented in C/C++. Originally, each of these benchmarks was implemented in CUDA and OpenCL parallel models. Recently, OpenACC implementation of the benchmarks has been added by a third-party \cite{Rodinia2013}. We include Dyadic Convolution and N-Body simulation from CUDA SDK and the remaining benchmarks from Rodinia.

{\bf OpenACC Compiler.} We use our in-house framework, IPMACC, for compiling OpenACC applications. IPMACC translates OpenACC to either CUDA or OpenCL and executes OpenACC application over CUDA or OpenCL runtime (e.g. NVIDIA GPUs or AMD GPUs). We validated the correctness of our framework by comparing the results of OpenACC benchmarks against the serial and CUDA version.

{\bf Performance evaluations.} We compile the OpenACC version of benchmarks by our framework and run it over CUDA runtime. We compare these to CUDA implementations available in CUDA SDK and Rodinia. In order to evaluate performance, we report the total time of kernel execution, kernel launch, and memory transfer between host and accelerator. We use \textit{nvprof} for measuring these times \cite{NVIDIA2014b}. For kernel execution and memory transfers time, we report the time that \textit{nvprof} reports after kernels/transfers completion. For kernel launch time, we report the time measured by \textit{nvprof} in calling \textit{cudaLaunch}, \textit{cudaSetupArgument}, and \textit{cudaConfigureCall} API procedures. Every reported number is the harmonic mean of 30 independent runs.

{\bf Platforms.} We perform the evaluations under a CUDA-capable accelerator. We use NVIDIA Tesla K20c as the accelerator. This system uses NVIDIA CUDA 6.0 \cite{NVIDIA2014a} as the CUDA implementation backend. The other specifications of this system are as follows: CPU: Intel® Xeon® CPU E5-2620, RAM: 16 GB, and operating system: Scientific Linux release 6.5 (Carbon) x86\_64. We use GNU GCC 4.4.7 for compiling C/C++ files.

\section{Performance Comparison}
In this section, we compare a set of OpenACC applications to their highly optimized CUDA version. Our goal is to identify OpenACC's programming limitations resulting in the performance gap between OpenACC and CUDA performance. See Methodology section for applications, compilers, and hardware setup.
\subsection{Performance Comparison}
\begin{figure}[t]
  \centering
  \begin{tikzpicture}
  \pgfplotsset{every axis legend/.append style={at={(0.5,1.03)},anchor=south}}
  \begin{axis}[ybar stacked,enlargelimits=0.01,
width=0.95\textwidth,height=6cm,ymin=0,ymax=2.6,
bar width=0.02\textwidth,area legend,
legend style={draw=none,font=\scriptsize}, legend columns=3,
tick align=outside,
ylabel=\textbf{Normalized Execution Time},
ymajorgrids,enlarge x limits=.1, enlarge y limits=0,
ytick={0, 0.2, 0.4, 0.6, 0.8, 1.0, 1.2, 1.4, 1.6, 1.8, 2.0, 2.2, 2.4, 2.6, 2.8},
yticklabels={0, 0.2, 0.4, 0.6, 0.8, 1.0, 1.2, 1.4, 1.6, 1.8, 2.0, 2.2, 2.4, 2.6, 2.8},
xtick=data, xticklabels from table={motivation2.dat}{bench},
xticklabel style={rotate=90,anchor=east}]
\addplot[draw=black,fill=red!90,postaction={pattern=north east lines}] table[x=num, y=memory] {motivation2.dat};
\addlegendentry{Memory transfer}
\addplot[draw=black,fill=blue!90,postaction={pattern=dots}] table[x=num, y=kernel] {motivation2.dat};
\addlegendentry{Kernel execution}
\addplot[draw=black,fill=yellow!40,postaction={pattern=north west lines}] table[x=num, y=launch] {motivation2.dat};
\addlegendentry{Launch overhead}
%\node[red,draw,circle] (red-2) at (11,1.6) {};
  \end{axis}
%\node[red,draw,circle,inner sep=4pt] at (red-2) {};
\node (title) at (0.08\textwidth,-1.6cm)  {\scriptsize Backprop};
\node (title) at (0.16\textwidth,-2cm) {\scriptsize BFS};
\node (title) at (0.24\textwidth,-1.6cm) {\scriptsize dyadic.};
\node (title) at (0.30\textwidth,-2cm) {\scriptsize Hotspot};
\node (title) at (0.38\textwidth,-1.6cm) {\scriptsize Matrix Mul.};
\node (title) at (0.44\textwidth,-2cm) {\scriptsize N-Body};
\node (title) at (0.52\textwidth,-1.6cm) {\scriptsize Nearest.};
\node (title) at (0.58\textwidth,-2cm) {\scriptsize Needle-Wunsch};
\node (title) at (0.66\textwidth,-1.6cm) {\scriptsize Pathfinder};
\node (title) at (0.72\textwidth,-2cm) {\scriptsize SRAD};
    \end{tikzpicture}
\caption{Comparing the execution time of OpenACC to highly-optimized CUDA implementations. Each bar shows the duration of time that the application spends on memory transfer, kernel execution, and kernel launch overhead.}\label{motiv}
\end{figure}
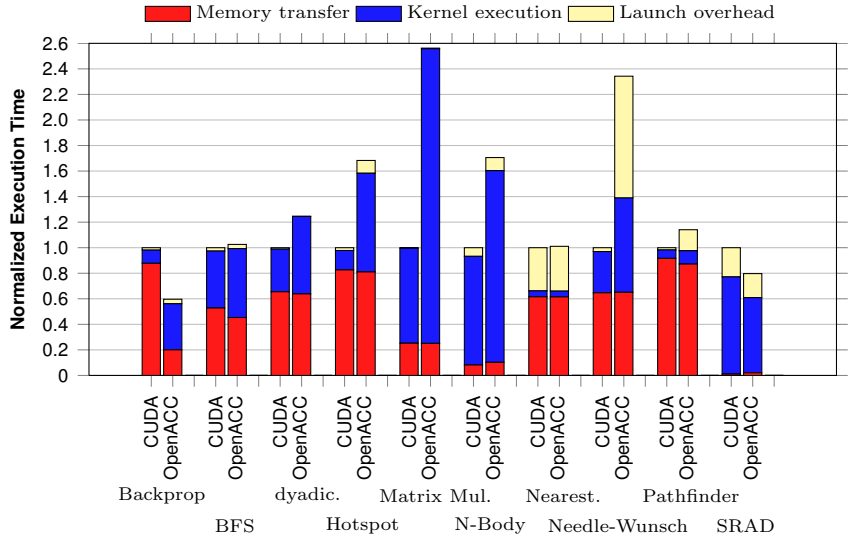

Figure \ref{motiv} reports the execution time for OpenACC applications, compared to their CUDA version. The figure reports the breakdown of time spent on the accelerator; kernel launch (launch), kernel execution (kernel), or memory transfer between host and accelerator (memory). Kernel launch time includes the time spent on setting kernel arguments and launching the kernel on the accelerator.
In most cases, CUDA's kernel launch/execution portion is shorter than OpenACC. Also, memory transfer times are comparable on both CUDA and OpenACC. There are exceptions where OpenACC memory transfers are faster (e.g. Backpro.) or kernel time of CUDA and OpenACC are equal (e.g. Nearest.). We investigate the differences between CUDA and OpenACC in the following sections.

\subsection{Investigating Performance Gap}
In this section, we discuss applications separately providing insight into why CUDA and OpenACC implementations presented in Figure \ref{motiv} have different kernel launch, kernel execution, and memory transfer times.

{\bf Back Propagation.} Back Propagation (Backpro.) is a machine-learning algorithm used to train the weights in a three-layer neural network. In both OpenACC and CUDA versions, there are six back-to-back serial operations where the output of each stage is fed to the immediate next stage as input. Each stage can be performed in parallel on the accelerator.

OpenACC implementation performs faster memory transfers and slower kernel launch/execution, compared to CUDA. This is explained by the fact that the OpenACC version executes all six stages on GPU, while the CUDA version alternates between CPU and GPU for execution. Alternating between CPU and GPU imposes extra memory transfer overhead. 

{\bf BFS.} {BFS visits all the nodes in the graph and computes the visiting cost of each node. Each node is visited only once. Parallel threads of a kernel visit the nodes belonging to the same graph depth concurrently and the algorithm traverses through the depth iteratively. The operation stops once there is no child to visit.}

Compared to the CUDA version, the OpenACC version of BFS spends less time on memory transfers. This can be explained by the fact that the OpenACC version performs data initializations on the GPU. However, the CUDA version initializes the inputs on the host and transfers the inputs to GPU. Compared to the CUDA version, OpenACC spends more time on kernel execution, since it forces a debilitating reduction on a global variable. {The global variable is a boolean indicating whether there remains more nodes to visit or not. CUDA avoids global reduction by initializing the variable to FALSE on the host and imposing a control-flow divergent in the kernel to guard the global variable from FALSE writes (allowing TRUE writes).}

{\bf Dyadic Convolution.} Dyadic Convolution (dyadic.) is an algebra operation calculating the XOR-convolution of two sequences. The OpenACC implementation parallelizes output calculations, where each thread calculates one output element. Although this implementation is fast to develop, it exhibits a high number of irregular memory accesses. To mitigate irregular memory accesses, the CUDA version uses Fast Walsch-Hadamard Transformation (FWHT) for implementing dyadic convolution (as described in \cite{Arndt2011}).

As reported in Figure \ref{motiv}, both OpenACC and CUDA versions spend almost the same amount of time on memory transfers. While the CUDA version launches several kernels, OpenACC launches only one kernel. This explains why the CUDA version imposes higher kernel launch overhead. In CUDA the kernels' execution time is 82\% faster than OpenACC. This is due to the fact that the CUDA version uses FWHT to mitigate irregular memory accesses. Although OpenACC can implement dyadic convolution using FWHT, the same FWHT algorithm used in CUDA cannot be implemented in OpenACC. CUDA FWHT uses shared memory to share intermediate writes locally between neighbor threads, which is not possible under OpenACC standard.

{\bf Hotspot.} Hotspot simulates chip characteristics to model the temperature of individual units. At every iteration, the algorithm reads the temperature and power consumption of each unit and calculates new temperatures. Although both OpenACC and CUDA spend the same amount of time on memory transfers, CUDA kernel is much faster.

In Hotspot, the temperature of each unit depends on its power consumption and neighbors' temperatures. CUDA kernel exploits this behavior to localize the communication and reduce global memory accesses as follows. In CUDA, threads of the same thread block calculate the temperature of neighbor units. The CUDA version locally updates the new temperature of neighbor units using the threads of the same thread block. This local communication reduces the number of kernel launches used to synchronize the temperature across all thread blocks, explaining why the CUDA version performs faster kernel launches and comes with shorter execution time. In OpenACC, unlike CUDA, the software-managed cache cannot be exploited for local communication. Hence, In OpenACC there are higher number of global synchronizations and kernel launches, which in turn harms performance.

\textbf{Matrix Multiplication.} Matrix Multiplication (Matrix Mul.) performs multiplication of two 1024x1024 matrices. Both CUDA and OpenACC implementations use output parallelization, calculating each element of the output matrix in parallel. CUDA version is different from OpenACC as it processes input matrices tile-by-tile. By processing in tiles, CUDA version fetches the input tiles in few well-coalesced accesses into software-managed cache and shares the tiles among the threads of the same thread block.

While kernel launch and memory transfer times are nearly the same across CUDA and OpenACC, CUDA kernel time is much lower than OpenACC. CUDA version takes advantage of software-managed cache in two ways. First, CUDA version merges the required data of the thread block and fetches them once, minimizing redundant memory accesses across thread of the same thread block. Second, software-managed cache removes cache conflict misses, since the replacement policy is controlled by the programmer. Under OpenACC, although the threads have very high spatial locality, parsing the matrix row-by-row at a time highly pollutes the cache, returning high number of conflict misses. Also having multiple thread blocks per SM exacerbates this effect.

{\bf N-Body simulation.} N-Body models a system of particles under the influence of gravity force. In each timestep, operations of O($N^{2}$) complexity are performed (for a system of N particles) to calculate forces between all pairs of particles. Inherently, there are many redundant memory reads, since the mass and position information of each particle is fetched by other particles N-1 times to calculate its interaction with other particles.

While both CUDA and OpenACC memory transfers take about the same time, CUDA kernels are much faster. The CUDA version tiles the computations to reduce redundant memory reads \cite{Nyland2007}. CUDA exploits shared memory to share the particles among all threads of a thread block. In OpenACC, however, the redundant memory accesses are not filtered out by the software-managed cache. As reported, redundant memory accesses can degrade performance significantly. 

{\bf Nearest Neighbor.} {Nearest Neighbor (Nearest.) finds the five closest points to a target position. The Euclidean distance between the target position and each of the points is calculated and the top five points with the lowest distance are returned.} OpenACC and CUDA versions both calculate Euclidean distances for each point in parallel. OpenACC and CUDA versions spend about the same time on kernel launch, kernel execution, and memory transfer. This is explained by the similarity of parallelization methods applied in both OpenACC and CUDA. 

{\bf Needleman-Wunsch.} {Needleman-Wunsch (Needle.) is a sequence alignment algorithm used in bioinformatics.} In either CUDA or OpenACC, traverses a 2D matrix and updates the costs. Upon updating a new cost, four memory locations are read and one location is written.

Although both CUDA and OpenACC versions spend the same amount of time on memory transfers, CUDA kernel launch/executions are much faster than OpenACC kernels. The CUDA version fetches a data chunk of \textit{costs} matrix into shared memory and traverses the matrix at the shared memory bandwidth. This mechanism comes with three advantages: i) filtering redundant global memory accesses by shared memory, ii) minimizing global communication by sharing intermediate results stored in the shared memory, iii) reducing the number of kernel launches and global communications. The fewer number of kernel launches explains why the launch time of CUDA is much less than OpenACC.

{\bf Pathfinder.} {In Pathfinder (Pathfin.) kernel, every working element iteratively finds the minimum of three consequent elements in an array. The CUDA version of Pathfinder performs two optimizations: i) finding the minimum by accessing the data from shared memory, and ii) sharing the updated minimum locally among neighbor threads for certain iterations and then reflecting the changes globally to other threads. Such local communications reduce the number of global synchronizations and kernel launches.}

However, OpenACC's API is not flexible enough to allow the programmer exploit the shared memory in a similar way. Therefore neighbor threads in the OpenACC version do not communicate via shared memory. Therefore, each thread fetches the same data multiple times and threads communicate only through global memory. Communication through global memory is implemented through consequent kernel launches. This explains why OpenACC imposes higher kernel launch overhead.

{\bf Speckle reducing anisotropic diffusion.} Speckle reducing anisotropic diffusion (SRAD) is an image processing benchmark performing noise reduction through partial differential equations iteratively. Compared to CUDA, the kernel time of OpenACC version is less. Three code blocks construct the computation iterative body of this benchmark: one reduction region and two data parallel computations. Our evaluation shows OpenACC version performs ~5\% slower than CUDA, upon executing two data parallel computations. However, OpenACC outperforms CUDA in executing the reduction portion. This is explained by the difference in reduction implementations. Our OpenACC framework performs the reduction in two levels: {reducing along threads of thread block on GPU and reducing along thread block on CPU. In the CUDA version, however, reduction is performed by multiple serial kernel launches, all on the GPU.} The OpenACC version spends less time on executing the kernel as part of the computation is carried on host. Meanwhile, performing two levels of reduction imposes the overhead of copying intermediate data from GPU to CPU. This explains why the OpenACC version spends slightly more time on memory transfers and less time on kernel launch/execution.

\section{Related Work}
Reyes et al. \cite{Reyes2012} introduce an open-source tool, named accULL, to execute OpenACC applications on accelerators. The tool consists of a source to source compiler and a runtime library. The compiler translates OpenACC notations to the runtime library routines. The runtime library routines are implemented in both CUDA and OpenCL. Tian et al. \cite{Tian2013} introduce an OpenACC implementation integrated in OpenUH \cite{Liao2007}. They evaluate the impact of mapping loop iterations over GPU parallel work-items.

Lee and Vetter \cite{Lee2014} introduce a framework for compiling, debugging, and profiling OpenACC applications. They also openarc directives allowing OpenACC programmer to map OpenACC arrays to CUDA memory spaces, including shared and texture memory spaces. They do not investigate the effectiveness of their proposal for these mappings. Based on the short introduction that they present, we believe their proposal for utilizing shared memory is different from ours in two ways. Firstly, while openarc directive needs programmer to separate shared memory array and corresponding global memory array in the code, fcw separates the arrays automatically, based on the information presented by the programmer. Secondly, while openarc directive allows fine-grained control to OpenACC programmer to perform fetch, synchronization, and writeback, fcw implicitly handles fetch, synchronization, and writeback. Based on these differences, we consider fcw as a high-level proposal for utilizing SMC and openarc as a low-level fine-grained control over SMC.

Nakao et al. \cite{Nakao2014} introduce XACC as an alternative to MPI+OpenACC programming model to harness the processing power of cluster of accelerators. XACC offers higher productivity since XACC abstractions reduce the programming efforts. Under small and medium problem sizes, XACC performs up to 2.7 times faster than MPI+OpenACC. This higher performance comes from the PEACH2 interface that XACC communicates through. PEACH2 performs faster than GPUDirect RDMA over InfiniBand under data transfer size of below 256KB. Increasing the problem size, XACC and MPI+OpenACC perform comparable, since the latency of PEACH2 and GPUDirect RDMA over InfiniBand would be equal.

%\bibliographystyle{abbrvnat}

% The bibliography should be embedded for final submission.

\end{document}